\documentclass[11pt,a4paper]{article}
\usepackage{jinstpub}
\usepackage{rotating}
\usepackage{graphicx}
\usepackage{upgreek}
\usepackage{caption}
\usepackage{subcaption}
\usepackage{array}

\begin{document}


\title{Characterization of CdTe Sensors with Schottky Contacts Coupled to Charge-Integrating Pixel Array Detectors for X-Ray Science}

\author[a,b]{Julian Becker,}
\author[a]{Mark W. Tate,}
\author[a]{Katherine S. Shanks,}
\author[a]{Hugh T. Philipp,}
\author[a,b]{Joel T. Weiss,}
\author[a]{Prafull Purohit,}
\author[b]{Darol Chamberlain,}
\author[b]{Jacob P. C. Ruff}
\author[a,b,1]{and Sol M. Gruner\note{Corresponding author}}

\affiliation[a]{Laboratory of Atomic and Solid State Physics, Cornell University, Ithaca, NY 14853,USA}
\affiliation[b]{Cornell High Energy Synchrotron Source (CHESS), Cornell University, Ithaca, NY 14853, USA}
\emailAdd{smg26@cornell.edu}


\abstract{Pixel Array Detectors (PADs) consist of an x-ray sensor layer bonded pixel-by-pixel to an underlying readout chip. This approach allows both the sensor and the custom pixel electronics to be tailored independently to best match the x-ray imaging requirements. Here we present characterizations of CdTe sensors hybridized with two different charge-integrating readout chips, the Keck PAD and the Mixed-Mode PAD (MM-PAD), both developed previously in our laboratory. The charge-integrating architecture of each of these PADs extends the instantaneous counting rate by many orders of magnitude beyond that obtainable with photon counting architectures. The Keck PAD chip consists of rapid, 8-frame, in-pixel storage elements with framing periods $<$150 ns. The second detector, the MM-PAD, has an extended dynamic range by utilizing an in-pixel overflow counter coupled with charge removal circuitry activated at each overflow. This allows the recording of signals from the single-photon level to tens of millions of x-rays/pixel/frame while framing at 1 kHz. Both detector chips consist of a 128$\times$128 pixel array with (150 $\upmu$m)$^2$ pixels.}

\maketitle

\section{Introduction}
The introduction of Pixel Array Detectors (PADs) and their subsequent use for x-ray sciences \cite{I1} revolutionized the way experiments are performed at synchrotrons in a variety of disciplines. By separating the processing layer, an Application Specific Integrated Circuit (ASIC), from the sensor layer, both can be optimized independently. To date the most common sensor material is silicon, not only because of its availability and low cost, but also because of its excellent quality. However, the stopping power of silicon for x-rays limits the ability to efficiently absorb (and ultimately detect) x-rays of energies above approximately 20~keV. 

\begin{figure*}[tb]
  \centering
  \includegraphics[width=0.8\textwidth]{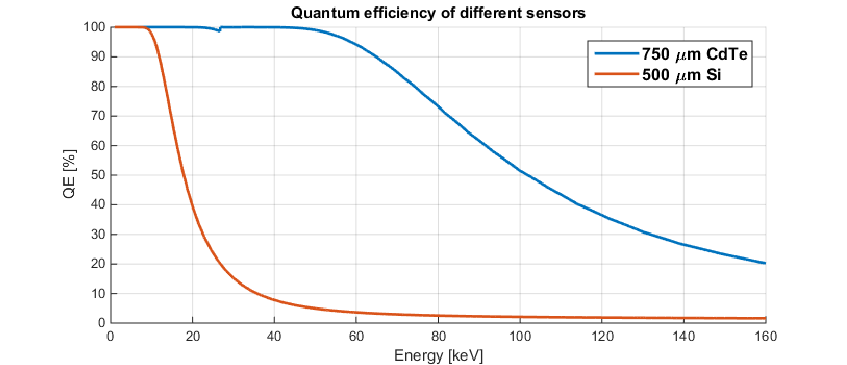}
  \caption{Quantum efficiency of 750~$\upmu$m CdTe and 500~$\upmu$m silicon sensors. }
  \label{QE}
\end{figure*}

For x-ray science applications at higher energies many different approaches exist \cite{detectors}; using Cadmium Telluride (CdTe) as a sensor material is one of them. The material and its quality have improved significantly in recent decades \cite{R1, R2} and to date there are several commercial cameras using CdTe for scientific imaging available on the market. 

We have chosen to develop versions of our detectors that allow experiments at higher x-ray energies. To this end we have replaced our standard 500~$\upmu$m thick silicon sensors with 750~$\upmu$m thick CdTe sensors. This change increases the quantum efficiency of the sensors in the 15 to 100~keV range and beyond, as shown in Figure \ref{QE}. 

Each of the possible layouts of CdTe sensors, i.e., ohmic or Schottky, has its particular advantages and disadvantages \cite{contacts, contacts2}. Both layouts differ only in the materials chosen for the contacts.

Ohmic layouts use the same metal for both electrodes and ohmic sensors effectively work as a photoresistor, i.e., the current flowing through the resistor is increased in the presence of x-rays. Commonly cited advantages of ohmic layouts are the ability to reverse the polarity of the device by reversing the bias voltage and a reduced susceptibility to polarization. A commonly cited disadvantage of ohmic material is the comparatively high leakage current, which can lead to increased noise in the readout system. 

Schottky layouts use different metals for anode and cathode, such that one of them becomes a barrier for either electrons or holes, which effectively turns the material into a photodiode. Reverse biased diodes are well suited for x-ray detection, as they feature a very low leakage current which allows fabrication of detectors with very low noise. However the polarity of the device has to be decided during the production of the sensor and cannot be changed afterwards. A commonly cited disadvantage of a Schottky layout is the tendency of the material to polarize quickly.  

We have chosen Schottky type material and hole collection for the sensor layout, as the architecture of the existing MM-PAD chip (explained below) is limited to hole collection and we require a low leakage current in order to maintain the performance of our detectors.

\subsection{Detector systems}
Cornell University has developed charge integrating detectors, two of which, the Keck PAD and the MM-PAD (detailed below) have been selected for hybridization with CdTe sensors. Both ASICs are fabricated in TSMC 0.25~$\upmu$m technology \cite{tsmc} and feature 128$\times$128 pixels of (150~$\upmu$m)$^2$ per chip.

Both detector chips can be tiled to create 2$\times$3 arrays, increasing the imaging area. The same custom built housing is used for both arrays, providing a thermally regulated vacuum environment, as well as support electronics.

\subsubsection{Keck PAD}
\begin{figure*}[tb]
  \centering
  \begin{subfigure}[t]{0.45\textwidth}
  	\includegraphics[width=\textwidth]{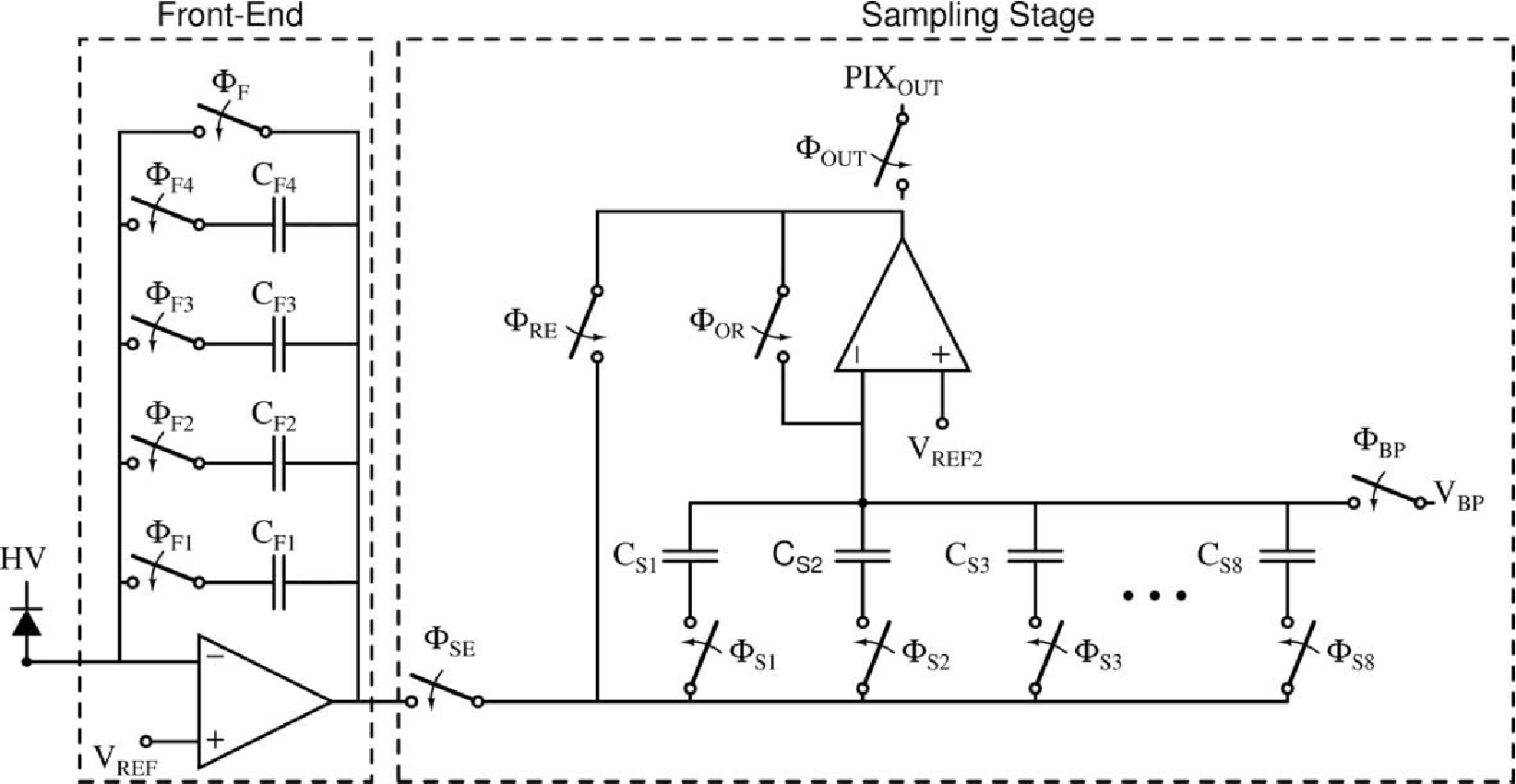}	
	\caption{Keck PAD block diagram}
	\label{Keck}
  \end{subfigure}
\quad
  \begin{subfigure}[t]{0.45\textwidth}
	\includegraphics[width=\textwidth]{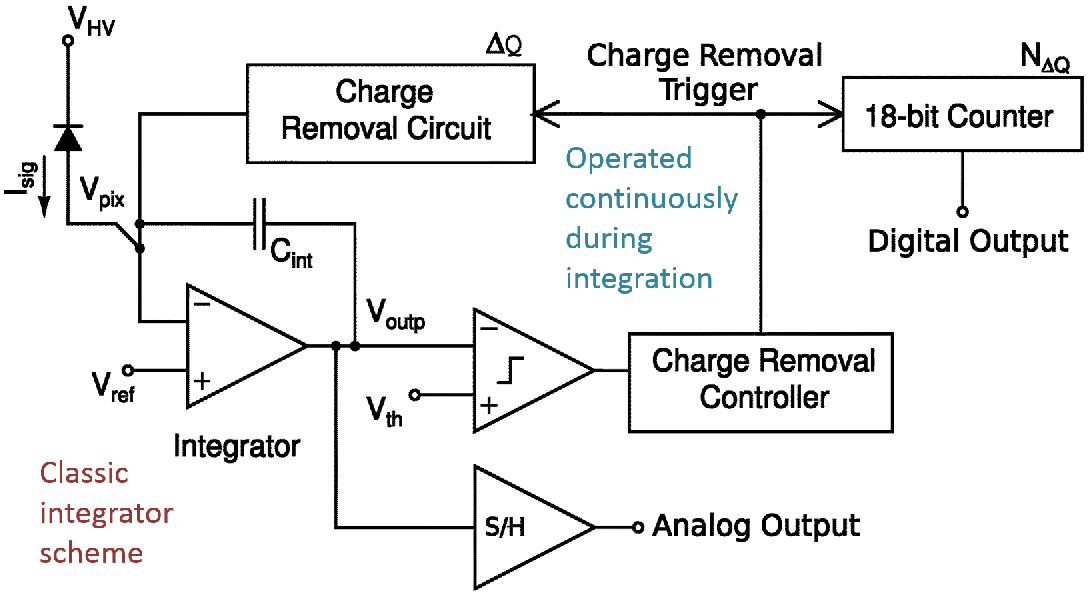}
	\caption{MM-PAD block diagram}
    	\label{MM}
  \end{subfigure}
  \caption{Schematic overview of the integrated circuitry of the detector systems hybridized with CdTe sensors.}
\end{figure*}

The Keck PAD \cite{K1,K2,K3,K4}, shown as a simplified schematic in Figure \ref{Keck}, 
shares operating principles with past detectors \cite{I1,O1}, like the analog integrating approach with in-pixel storage, 
which in turn inspired current day burst mode imagers like the AGIPD \cite{A1,A2}.
Burst imaging at up to 10 MHz is possible (100~ns pulse separation) and the front-end charge-to-voltage conversion gain is set by the configuration of switches $\Phi_{\rm F1}$-$\Phi_{\rm F4}$, making the gain adjustable by up to a factor of 6.5. 


In addition, the front-end capacitors ($C_{\rm F1}$-$C_{\rm F4}$) may also be re-addressed to add signal without reading out the device. 
In essence this feature allows the detector to work as a lock-in amplifier for x-rays.

\subsubsection{Mixed-Mode PAD (MM-PAD)}
The MM-PAD \cite{M1,M2,M3} uses an integrating approach with counting features, as shown in Figure \ref{MM}.
Charge is accumulated on the integration capacitor until the output voltage of the front-end
integration stage passes a programmed threshold, generally equivalent to a few hundred 8~keV photons. 

When the output voltage reaches this level, an in-pixel circuit removes a fixed amount of charge, typically also equivalent to a few hundred 8~keV photons, from the integration capacitor. This charge-removal process can occur concurrently with the arrival of charge from stopped x-rays, i.e., the process incurs no dead-time. 

An in-pixel digital counter records the number of times the charge removal circuit is triggered. At the end of the integration period, the in-pixel digital counter is read out, as well as the analog output of the front-end integrating amplifier. The analog output is digitized with off-chip electronics and is combined with the digital counter output to measure the total charge produced by x-rays absorbed in the sensor.

The MM-PAD achieves single x-ray sensitivity \cite{M4} and spans a dynamic range of $>$~4$\times$10$^7$ x-rays/pixel/frame (at 8~keV) while framing at $>$~1 kHz. This has proven to be very useful for coherent x-ray imaging \cite{M5}. 

\section{Material and Methods}

The new CdTe hybrids were characterized with a range of tests using lab sources as well as synchrotron radiation from the Cornell High Energy Synchrotron Source (CHESS). 

\subsection{Radiation sources}
Sensor response was measured for a broad range of photon energies using the following sources:

\begin{enumerate}
	\item A 50~W silver (Ag) anode x-ray tube operated at an acceleration voltage of up to 47~kV. The $K_\alpha$ and $K_\beta$ lines of silver are at 22.16 and 24.94~keV, respectively, with a $K_\beta/K_\alpha$ ratio of 0.217 \cite{fl_ratio}. A filter of 1~mm aluminum was optionally used to attenuate low energy x-rays.
	\item An americium-241 isotope source with a nominal activity of 100~$\upmu$Ci. The source provided alpha particles, as well as the Am-241 gamma spectrum with a prominent x-ray emission line at 59.5~keV.
	\item The beamline A2 at CHESS. For the experiments presented here, the white undulator beam was filtered by 1.5 mm of highly-ordered pyrolytic graphite, 0.76 mm of water-cooled aluminum, and 3.5 mm of water-cooled copper. This resulted in a spectrum peaked around 90 keV, with broad bandpass dE/E~$\approx$~0.5.
\end{enumerate}

\subsection{Experimental setups}

%
%
%

The x-ray tube was either mounted with a graphite monochromator crystal interposed to select the x-ray energy incident on the detector or as close as possible ($\approx$~5~cm) to the detector to maximize the incident flux.  For certain measurements an array of pinholes (50~$\upmu$m hole diameter on a square array of 0.292~mm pitch in 75~$\upmu$m thick tungsten) or a 200~$\upmu$m thick tungsten knife edge was interposed. In addition a lead line pair mask with several line periods was used as an imaging example.

For tests involving the americium-241 source a special sample holder replaced the vacuum window of the detector housing. In this case the source was located approximately 8 cm from the sensor in the vacuum, thereby allowing the alpha particles emitted by the source to reach the detector without absorption by the window or air.

The A2 beamline is fed by a CHESS Compact Undulator \cite{c1, c2}, sourced by 5.3 GeV electrons. The resulting x-ray beams deliver high flux to very high energies, more than 200~keV at the time of our experiments. In the experiment we used a `blue beam' configuration, where the highest-energy tail of the polychromatic spectrum can be safely delivered to a sample after being hardened by a series of high-pass filters. We used a 4-circle diffractometer to perform the high-energy transmission Laue studies of single crystals. The detector was mounted on a 2 theta arm facing the sample. For some experiments, instead of using a sample in the holder, a thick copper plate was placed upstream of the sample holder to create a diffuse scattering pattern. Additionally thin elemental sheets could be installed to shadow parts of the detector.\footnote{We used Ho, Yb and Re foils with K-edges at 55.62, 61.33 and 71.69~keV, respectively.} This way the selective absorption above and below the element's K-edge energy could be utilized to calibrate the gain of the detector. Installation of a piece of single crystal silicon in the sample holder, with suitable adjustment of incident and exit beam angles, allowed selection of a monochromatized beam for calibration purposes. 

\subsection{Data analysis}
The raw data stream of the MM-PAD consists of a digital value and the digitized remainder of the analog signal. For data analysis the digital value is multiplied by a scale factor and added to the digitized remainder to form an Analog Digitized Unit (ADU) \cite{mm_gain}. Data processing is done on these ADUs. The scale factor accounts for the amount of charge removed by a single charge removal operation. It should be noted that the scale factor was adjusted for different temperatures.
 
Data from the Keck detector is transferred off-chip as an analog value and digitized in the readout electronics. No preprocessing of the ADUs is necessary for the Keck system.

Although the CdTe sensor was fabricated with a guard ring to minimize edge effects, increased signal was found within the first few pixels from the edge in both detectors and attributed to residual edge effects. To fully eliminate edge effects in the data analysis a rim of 15 pixels from the edge was excluded from all analyses.

\subsubsection{Dark frames and drift corrections}

Both the MM-PAD and the Keck PAD are integrating detector systems. For integrating systems all frames taken, including dark frames, are subject to an offset determined by the working points of the readout electronics and the amount of current integrated during the integration time.
To remove this offset each frame had an average dark frame subtracted. This dark frame was the average of 20 frames using an identical integration time.

In the presence of a very weak signal it is possible to additionally correct for slow drifts in the background value due to the polarization of the CdTe material with time (investigated below). This drift correction relies on the fact that for any given number of weak frames the absence of a photon is the most probable event, thus `zero' is the most common reading. Finding the most common value in a series of consecutive images that is short compared to the typical polarization time yields the necessary correction for this series. For the studies presented here about 100 to 250 consecutive frames were enough to determine the offset with sufficient accuracy and, given our typical integration time of several milliseconds, this means that this correction is sufficient to compensate for polarization on the timescale of seconds or longer, which is well suited to compensate the observed effects on the timescale of several minutes to hours.

\subsubsection{Flat field corrections}

\begin{figure*}[tb!]
  \centering
  \begin{subfigure}[t]{0.45\textwidth}
	\includegraphics[width=\textwidth]{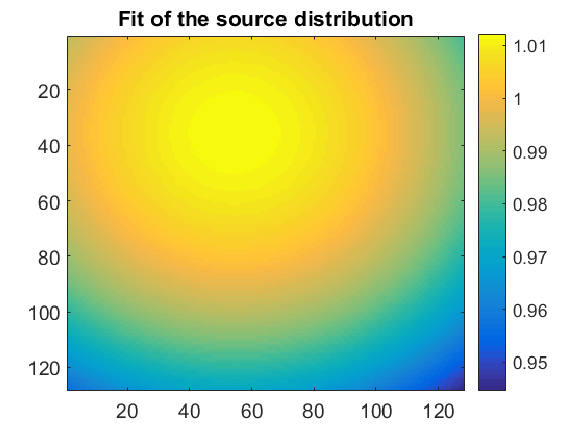}
	\caption{Relative distribution of the source intensity determined from a 2-dimensional quadratic polynomial surface fit to the data.}
	\label{source_2d}
  \end{subfigure}
\quad
  \begin{subfigure}[t]{0.45\textwidth}
	\includegraphics[width=\textwidth]{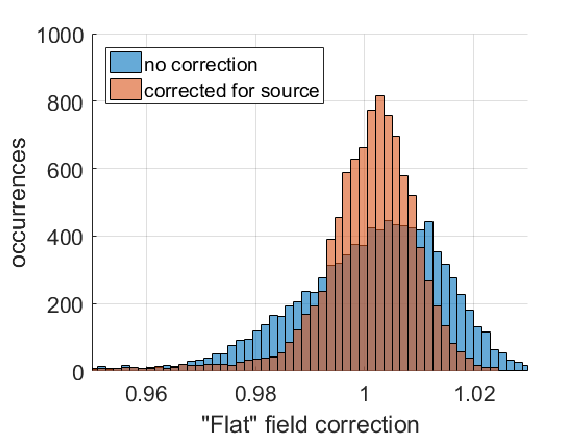}
	\caption{Histogram of a given flat field with and without correction for the source distribution. Structure remaining after the correction is a feature of the sensor material.}
	\label{source_res}
  \end{subfigure}

  \caption{Typical distribution of measurement values of the MM-PAD when the x-ray tube is close to the detector.}
  \label{source}
\end{figure*}

Both detector systems show fixed pattern distortions in the image, which can be corrected by flat fielding. These distortions can either be random (e.g., due to process variations in the ASIC affecting the gain of the pixel) or systematic (e.g., due to lateral displacement of charge carriers during the charge collection process) and inspection of the flat field data reveals information about possible structural effects of the sensor material (shown in detail below). The flat field correction for each pixel can be calculated from the ratio of its response to uniform illumination, $I_{pix}$, and the average response of all pixels to the same uniform illumination, $I_{avg}$.

To increase the dose rate, some flat field illuminations have been acquired with the x-ray tube in close proximity (approximately 5~cm). At this distance the size of the detector is no longer small compared to the distance, $r$, from the tube and a geometric correction needs to be applied to recover a uniform illumination. 

Idealizing the tube as a point source, the intensity in each pixel is approximately proportional to $1/r^2$, with $r=\sqrt{r_0^2 + p^2\left( (x-x_0)^2 + (y-y_0)^2 \right)}$, where $r_0$ is the smallest distance between source point and detector plane, $p$ is the linear pixel size, $x$ and $y$ are the pixel coordinates, and $x_0$ and $y_0$ are the points where the perpendicular intersects the detector plane\footnote{This derivation also holds when the foot of the perpendicular ($x_0$,$y_0$) falls outside of the actual detector {\it surface} onto an extended detector {\it plane}. Also $x_0$ and $y_0$ are not necessarily integer values.}.

Due to this geometric effect, the illumination is not truly uniform. To compensate for this the background subtracted data was fit to a 2-dimensional quadratic polynomial function. The flat field response was then determined by dividing the background subtracted data by the fit value for each pixel. Fit results of the 2-dimensional quadratic polynomial function fitted to a typical data frame when the x-ray tube in close proximity are shown in Figure \ref{source_2d}. The data points are displaced systematically between +1\% and -5\% around the average value of 1. Comparing the histogram of a flat field with and without correcting for the geometric distortion (Figure \ref{source_res}) underlines the importance of this correction.

\subsubsection{Distributions and statistical methods}

\begin{figure*}[tb!]
  \centering
  \begin{subfigure}[t]{0.45\textwidth}
	\includegraphics[width=\textwidth]{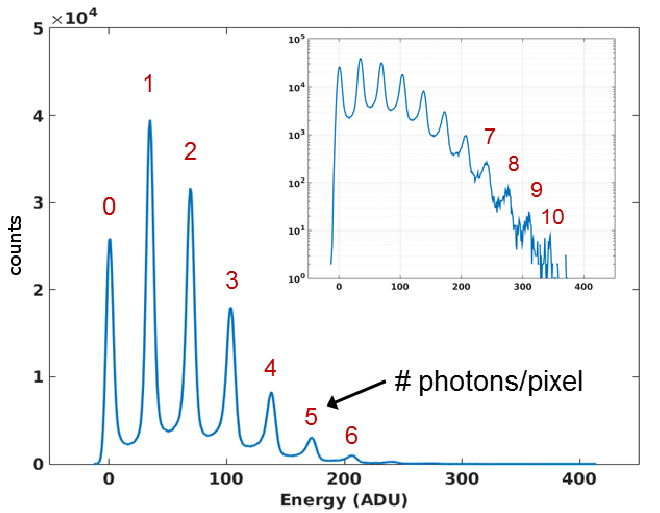}
	\caption{Histogram of pixel measurement values on the $K_\alpha$ line of the silver tube acquired using a pin hole array. The multi-modal distribution is readily observable and, using a logarithmic scale (insert), up to 10 photons (11 modes) can be counted.}
	\label{p_ag}
  \end{subfigure}
\quad
  \begin{subfigure}[t]{0.50\textwidth}
	\includegraphics[width=\textwidth]{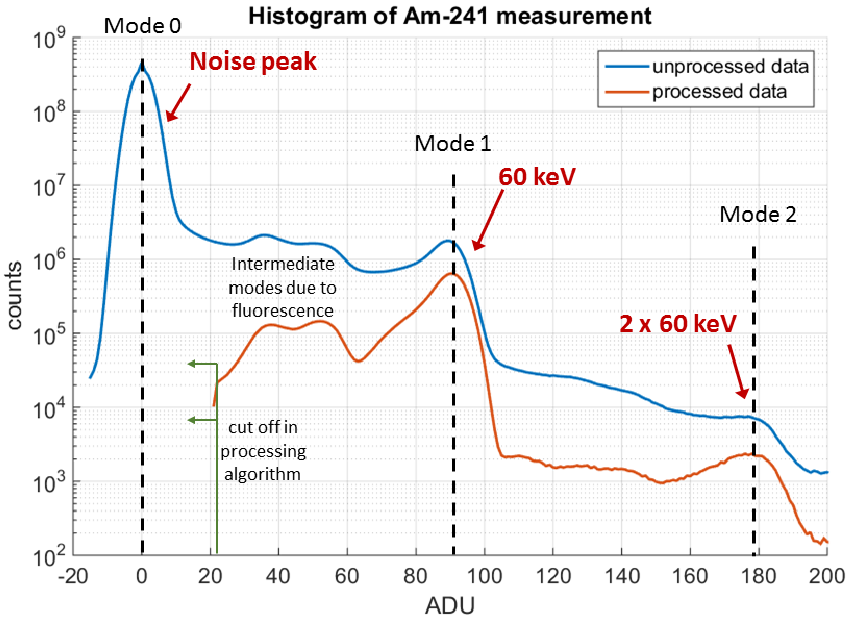}
	\caption{Histogram of the pixel measurement values using an Am-241 source (blue) and processed data (orange). Several distortions due to fluorescence and charge sharing can be observed on top of the expected multi-modal distribution.}
	\label{p_am}
  \end{subfigure}

  \caption{Typical distribution of pixel measurement values (spectrum) from the MM-PAD with CdTe for low flux measurements.}
  \label{p_low}
\end{figure*}

Much of the information presented here is derived from statistical analysis of various data sets. In some cases, e.g., the determination of the background offsets, the distribution of measurement values is unimodal and can be described well by a Gaussian normal distribution. The mean background value can thus be determined by a simple average of the measurement values, the (read-) noise of the system can be measured by determining the standard deviation of the measurement series. Given enough measurement points the estimates will be very close to the true values.

The situation becomes more complicated once the sensor is illuminated. Details on the way photons are absorbed and detected, the associated complications, and their probabilities are available in literature \cite{horus, sps}. For the sake of simplicity we assume here that the number of photons is a non-negative integer number and follows a Poisson distribution. Hence the probability, P, to have N photons is $P(N)=\frac{\lambda^N e^{-\lambda}}{N!}$, with $\lambda$ being the mean number of photons per illumination time. For polychromatic sources such as x-ray tubes the mean number of photons, and the energy of the detected photons, is a function of the source spectrum.

The signal of each pixel is generally proportional to the sum of the energies of all photons absorbed in that pixel. However subsequent interactions (e.g., fluorescent photon escape) may modify the response, and each signal is subject to read noise.  

For small average numbers of detected photons and a large enough signal-to-noise ratio the distribution of measurement values becomes multi-modal, and the location of peaks correspond to the product of the number and energy of detected photons. An example of such a distribution is shown in Figure \ref{p_ag}, where peaks from 0-10 photons/pixel/frame can be seen. For this measurement an array of 50~$\upmu$m pinholes was used to reduce charge sharing between pixels. The width of the peaks in Figure \ref{p_ag} is due to the read noise of the system.


In addition to read noise the spectral distribution is generally affected by charge sharing and Cd and Te fluorescence; an example of this is shown in Figure \ref{p_am}. Fluorescence can create intermediate modes, which can complicate the identification of mode numbers. This is evident in Figure \ref{p_am}, where a Cd K$_\alpha$ peak is observed around 36 ADU and a corresponding escape peak at around 54 ADU.

Once the number of photons per illumination time becomes large enough, spectral features smear out and the mean measurement value is roughly proportional to the average number of photons multiplied by their respective energy. The spread of the measurement values is determined by both the read noise and the inherent spread of the Poisson distribution.

Wherever possible we performed statistical analysis on a per-pixel basis; however, for certain measurements the number of events was too low for reliable parameter extraction. In this case the analysis was performed for the entire data set, treating all pixels as equal. The pixel-to-pixel variations, usually around 1\%, introduce an additional systematic uncertainty in the result.

\subsubsection{Event reconstruction by image processing}

\begin{figure*}[tb!]
  \centering
  \begin{subfigure}[t]{0.45\textwidth}
	\includegraphics[width=\textwidth]{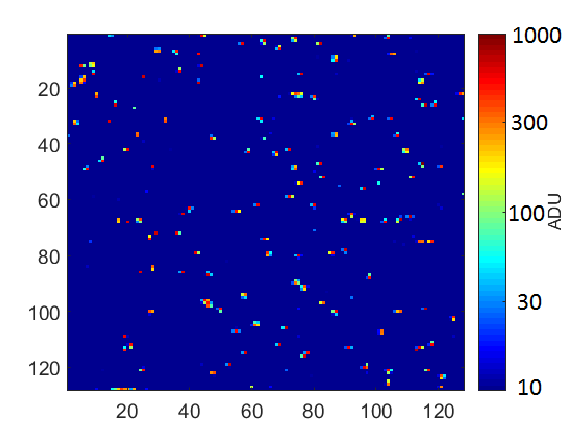}
	\caption{Single data frame acquired with 1~ms exposure using the Am-241 before image processing. Most alpha particle events are isolated clusters of 2 - 4 pixels, very few events overlap.}
	\label{a1}
  \end{subfigure}
\quad
  \begin{subfigure}[t]{0.45\textwidth}
	\includegraphics[width=\textwidth]{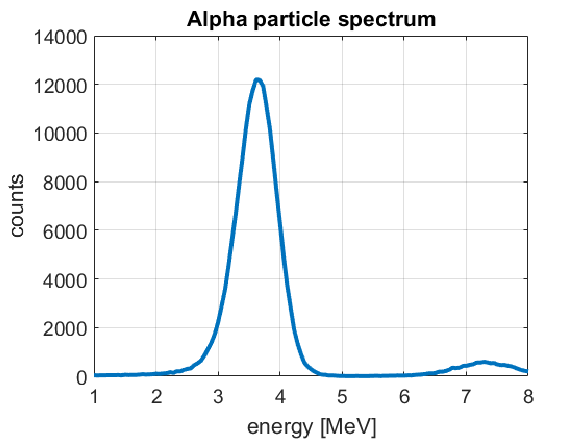}
	\caption{Spectrum of the alpha particles after image processing. An energy of approximately 3.8~MeV is measured due to energy losses along the trajectory of the alpha particle.}
	\label{a2}
  \end{subfigure}

  \caption{Measurements of alpha particles using the Keck PAD with 600~V bias at 0~C.}
  \label{alpha}
\end{figure*}

The signal generated by a single photon might be split among multiple pixels. Several mechanisms for this exist, the sharing of the charge cloud generated by a photon and fluorescence photons exciting neighboring pixels being the most important ones for this study. There is about 80\% probability to generate a fluorescence photon for each incident photon above the Te K-edge (about 31.8~keV) and those photons have a mean travel distance before absorption of about 128~$\upmu$m for the Cd fluorescence and 64~$\upmu$m for Te fluorescence \cite{david}. This means there is a non-negligible chance for each individual photon event to deposit of a significant fraction of the original photon energy (23.2~keV if a Cd K$_\alpha$ photon is produced) in a neighboring pixel.

In the case of sparse data, i.e., one photon or less in any given 3$\times$3 cluster of pixels, the original photon energy can be reconstructed by means of image processing. An algorithm was developed in MATLAB \cite{matlab} that would find these split events, sum them up and assign the sum to the pixel contributing the most to the sum. This process removes a large fraction of the split charge events and improves spectral clarity; however, it also increases the noise associated with the individual peaks (i.e., their FHWM) in the reconstructed energy spectrum by a factor proportional to $\sqrt{N}$, where $N$ is the number of pixels summed together in the reconstruction. A comparison of raw and processed data is shown in Figure \ref{p_am}. There is a bit of trial and error involved in the definition of the best threshold and exclusion criteria. Given the assumption that sigma is the average noise level of a pixel, the following criteria have been used in our analyses: A) require a seed pixel with a signal above 5 sigma, B) add neighbor to seed pixel if its value is above 3 sigma, C) require that only 1 pixel in the 3$\times$3 pixel area surrounding the seed is above 3 sigma, otherwise discard the event.

Figure \ref{a1} shows the case of alpha particles, where large clusters of signal dominate. We defined a signal cluster as a group of adjacent pixels that have signal values significantly exceeding the background reading. In order to evaluate the total energy deposited by each alpha particle all the pixels in each cluster are summed and the result is shown in Figure \ref{a2}. The algorithm used to do this is similar to the one described above, but iteratively expands the number of pixels belonging to a cluster as long as there are still adjacent pixels above the threshold. Note that the measured energy is less than 5.4~MeV, the decay energy of alpha particles from Am-241. Alpha particles are emitted from the source with approximately 4.7~MeV energy due to the encapsulation and another 0.9~MeV energy are lost in the Indium electrode and due to residual gas collisions in the vacuum chamber. 

\subsubsection{Numerical simulations of expected results}

\begin{table*}
\centering	
\begin{tabular}{>{\centering}m{0.1\textwidth}|>{\centering}m{0.2\textwidth}|>{\centering}m{0.2\textwidth}|>{\centering}m{0.2\textwidth}|>{\centering}m{0.2\textwidth}}
				& \textbf{electron mobility $\mu_e$}	& \textbf{hole mobility $\mu_h$}	& \textbf{elec. trapping time $\tau_e$}	& \textbf{hole trapping time $\tau_h$} 
\tabularnewline
\hline
\textbf{value}		& 1100 cm$^2$/Vs \cite{mob}	& 88 cm$^2$/Vs \cite{mob,mob2}	& 3~$\upmu$s \cite{acrorad}	& 2-4~$\upmu$s \cite{acrorad} 
\tabularnewline
\hline
\textbf{comments}		& temperature dependence reported; \\small sensitivity in measurements due to layout (hole collection)			& approximately constant in investigated temperature range					& $\gg$ than drift time; \\almost negligible in our case					&  determines charge collection efficiency in our case
\end{tabular}
\caption{Material constants used for the drift-diffusion simulations.}
\label{sim}
\end{table*}

Expected sensor response was simulated using custom MATLAB code. The code implements the algorithms described in \cite{sim1, sim2} to perform one dimensional drift and diffusion simulations, from which important parameters like the charge collection efficiency (i.e., the ratio of collected to generated charge) were extracted.

For these simulations we assumed the electric field in the sensor is linear, as is the case in filly depleted silicon sensors, and a bias voltage of 50~V is required to fully deplete the sensor. Further assumptions are constant mobility and trapping time throughout the sensor for electrons and holes in the investigated temperature range. Hole trapping times were adjusted for different temperatures, while all other parameters were unchanged with temperature. The chosen values are displayed in Table \ref{sim}.

It should be noted that while the published mobilities of electrons depend on the manufacturer of the CdTe (1100~cm$^2$/Vs \cite{mob} vs. 880~cm$^2$/Vs \cite{mob2}) this is not the case for the hole mobility (88~cm$^2$/Vs \cite{mob} vs. 90~cm$^2$/Vs \cite{mob2}). Due to the hole collecting readout and the small pixel effect \cite{spe} simulation results almost exclusively depend on the mobility and trapping time of the holes. In fact simulations that were run with the `wrong' set of parameters reproduced the results of the `right' set within 0.1\% .

\section{CdTe Material}
\begin{figure*}[tb!]
  \centering
  \begin{subfigure}[c]{0.45\textwidth}
	\includegraphics[width=\textwidth]{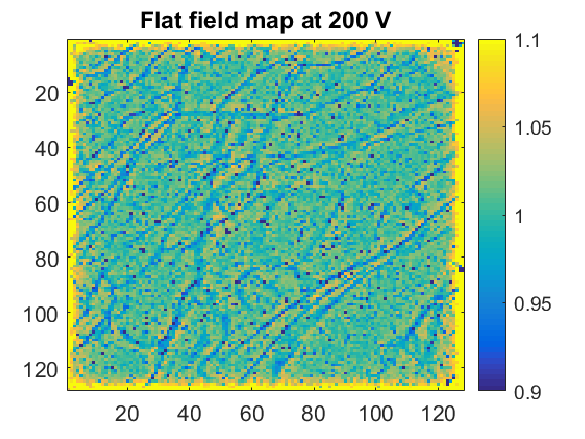}
  \end{subfigure}
\quad
  \begin{subfigure}[c]{0.45\textwidth}
	\includegraphics[width=\textwidth]{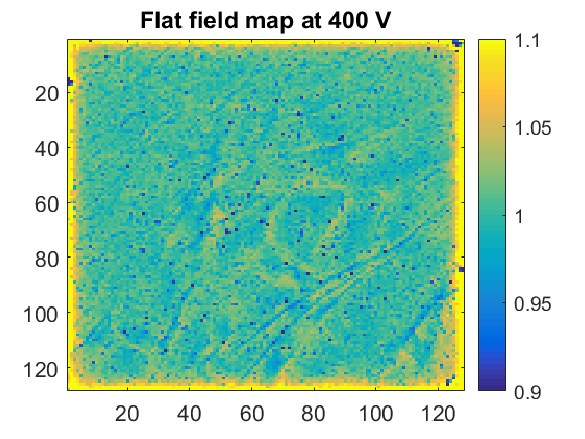}
  \end{subfigure}

  \begin{subfigure}[c]{0.45\textwidth}
	\includegraphics[width=\textwidth]{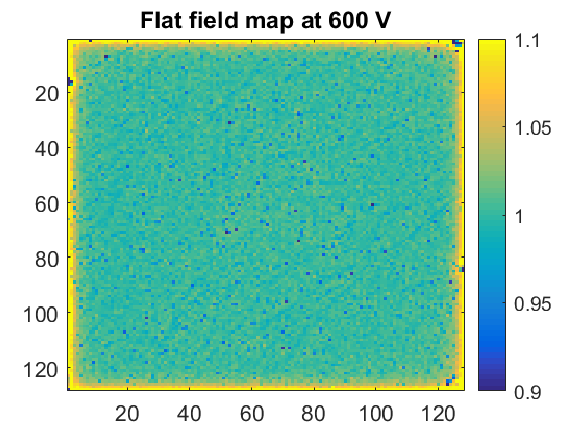}
  \end{subfigure}
  \quad
  \begin{subfigure}[c]{0.45\textwidth}
	\includegraphics[width=\textwidth]{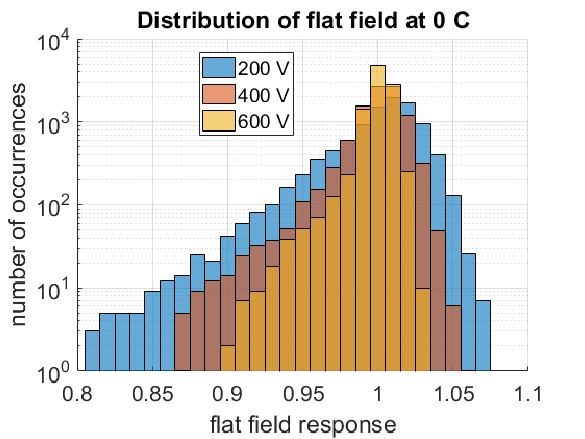}
  \end{subfigure}
  \caption{Flat field responses as a function of sensor bias voltage determined with an aluminum filtered silver tube at 47~kV using the MM-PAD detector in close proximity to the tube and correcting for the source distribution.}
  \label{flat_field}
\end{figure*}

The CdTe material used for our sensor investigation is 750~$\upmu$m thick and was produced by Acrorad Co., Ltd., Japan \cite{acrorad}. The sensors are In/Pt Schottky type with a pixellated Pt contact. In this way we collect holes. Fabricating the sensors with ohmic contacts (Pt on both sides) or pixellating the other contact was considered, but abandoned due to our requirements of hole collection and low leakage currents. The pixel matrix is surrounded by a single 120~$\upmu$m wide guard ring that is contacted and biased at the pixel potential to collect current generated outside of the area of the pixel matrix.

Bump deposition and flip-chip bonding were performed by Oy Ajat Ltd., Finland \cite{ayat}. Placement and gluing of the modules onto heat sinks were performed in-house and wire-bonding of the ASIC was done by Majelac Technologies LLC, USA \cite{majelac}. The HV wire used to bias the sensor was attached in-house using a conductive silver-epoxy glue. All glues were cured at room temperature.

Inspection of flat field data (shown in Figure \ref{flat_field}) collected with the MM-PAD system does not reveal any insensitive, grain-like defects in the material. These defects have been reported in the past by many groups \cite{grains1, grains2, grains3, greiffenberg}. Their lack might be an indication of increased quality. A network of lines is visible in the flat field at low bias voltage and reduces in contrast with higher voltages. Increasing the bias reduces the low response tail and narrows the distribution of the flat field correction (lower right hand plot in Figure \ref{flat_field}), which is indicative of a more uniform response. The detector was reset between the measurements. Details on the biasing and the reset scheme are provided in a later section. Other than the network of lines no significant distortions are observed. However, it should be mentioned that the network of lines increases in contrast once the material starts to polarize (investigated in a later section).


\subsection{Operating conditions}
All chips were mounted on temperature controlled heat sinks in an evacuated housing. The operating temperature could be controlled to $\pm$~0.1~C by a thermoelectric element, which in turn was cooled by chilled water. The operating temperature was adjustable in the range between +30 and -30~C. The high voltage bias was supplied by an external voltage source and was adjustable up to 600~V, the limit being set by the design of our system rather than the material itself.

Whenever possible the material was depolarized (reset), as explained in the following section, before a measurement was done. 

\subsection{Reset procedure}

The response of CdTe is known to vary as a function of time and exposure. These changes are usually summarily called `polarization' effects and are commonly suspected to be caused by trapping of charge carriers. Removing residual effects from previous illuminations is imperative for a good imaging system. Several methods of resetting the detector were tested and it was found that a simple bias refresh, a common procedure in photon counting detectors, was insufficient to clear the effects investigated in this work and presented in detail below. The reset procedure was optimized using the lateral displacement as an indicator of polarization.\footnote{Briefly, a lateral displacement field is observed after differential illumination and is seen at doses an order of magnitude lower than other polarization effects. The effect is explained in more detail a later section.}

Applying a forward bias to the diode essentially floods the sensitive region with opposite polarity charge carriers, in our case electrons. We speculate that given sufficient charge (current~$\times$~time) we can force a recombination of the trapped holes with these `clearing' electrons, thereby `resetting' the sensor. Investigating different clearing currents by increasing the forward bias of the diode we found that higher currents could clear a given amount of polarization in less time. However exact quantification of this proved difficult. It was found that for most cases presented in this work applying a forward bias of 5~V for 1 minute was sufficient to clear the accumulated polarization.\footnote{Since the input node of each pixel is actively being held at approximately 1.5~V (virtual ground of the preamplifier), the actual forward bias across the diode is 5~V~$+$~1.5~V~$=$~6.5~V.} After several minutes of heavy dosing at the CHESS beamline A2 with a high flux ($>$ 10$^{11}$ photons/mm$^2$/s) of 75~keV photons, we increased the reset time to 10 minutes for a complete reset, as a 1 minute reset was insufficient to completely clear the sensor.



A typical reset cycle is as follows: 1)~ramp the bias down to -5~V, 2)~wait for at least 1~minute, 3)~ramp HV up to the desired bias voltage, 4)~wait at least 3~minutes for the dark current to stabilize and 5)~take a new set of background images. The suggested wait time at point 4 depends on the type of exposures that are desired in the next measurement set. For short exposures of 1~$\upmu$s or less a wait time of a minute is sufficient, as the dark current does not contribute much to the measured signal. For exposures of 10~ms or more, a longer wait time is recommended.

In order to increase the effectiveness of the reset the wait time at step 2 can be increased. Further improvement was observed when the temperature of the detector is raised to +30~C during this waiting time. It should be noted that in our case, when temperature was increased during the waiting time, reaching a stable operating temperature again after the reset cycle took longer than the entire reset cycle.

From these results and the observation of the effects presented in the following, we concluded that the detector should undergo a reset cycle at least once every few hours,\footnote{It is well known that Schottky type CdTe sensors polarize even without being illuminated. The impact of this effect for our systems  is investigated in a later section.} after being exposed to approximately 10$^{10}$ photons$/$mm$^2$, and preferably in between individual measurement sets as well, in order to minimize the influence of polarization effects on the measurement results.

\subsection{Charge collection efficiency and gain calibration}

\begin{figure*}[tb!]
  \centering
  \begin{subfigure}[t]{0.45\textwidth}
  	\includegraphics[width=\textwidth]{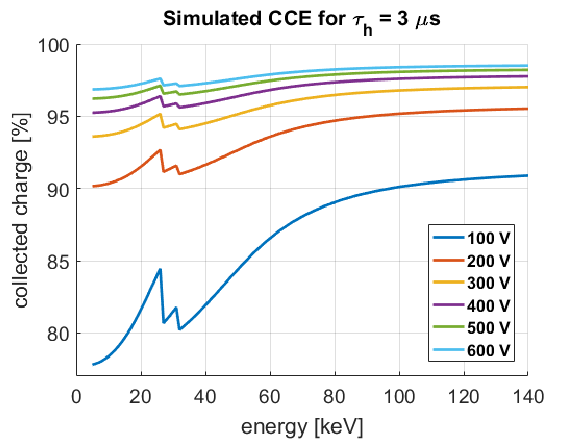}
	\caption{Simulated CCE as a function of photon energy for different voltages assuming constant mobility and a hole trapping time of 3~$\upmu$s.}
	\label{cce}
  \end{subfigure}
\quad
  \begin{subfigure}[t]{0.45\textwidth}
	\includegraphics[width=\textwidth]{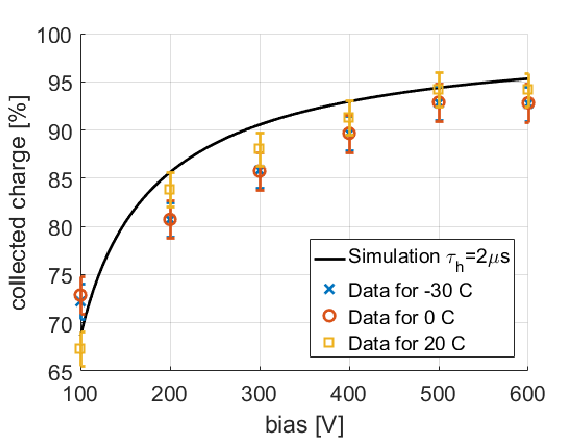}
	\caption{Charge collection efficiency of alpha particles as a function of voltage determined using the Keck PAD at different temperatures.}
	\label{cce_alpha}
  \end{subfigure}
  \caption{Simulation of the charge collection efficiency for photons as a function of energy and measurements of the CCE for alpha particles as a function of voltage.}
\end{figure*}

\begin{figure*}[tb!]
  \centering
 	\includegraphics[width=0.65\textwidth]{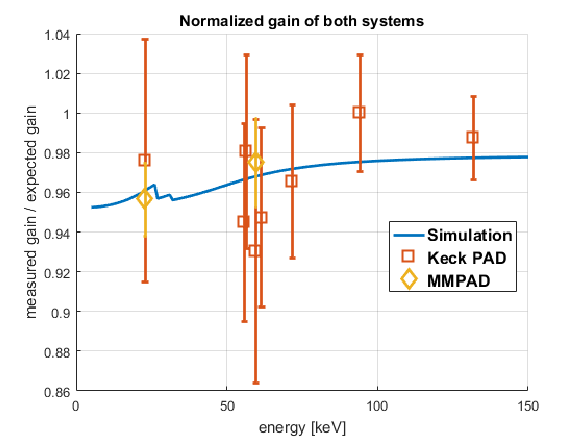}
	\caption{Measured gain of Keck PAD and MM-PAD for different x-ray energies in relation to the expected gain at complete charge collection efficiency. All measurement were done at 400~V bias and 0~C.}
	\label{gain_erg}
\end{figure*}

Owing to the fact that the charge carrier trapping time is only one order of magnitude larger than the typical drift time of charges through the detector, an incomplete charge collection is expected and the charge collection efficiency (CCE) is dependent on the energy of the incident photons as shown by simulations (Figure \ref{cce}). Figure \ref{cce_alpha} shows the measured CCE for alpha particles as a function of voltage. The figure also includes a numeric CCE simulation, assuming a hole trapping time of 2~$\upmu$s.


To establish a response curve for photons of different energies, histograms of pixel response at low fluence were measured. 
Figure \ref{gain_erg} shows the normalized gain for different energies. The calibrated energies include Ag K photons from the x-ray tube, 59.5~keV photons from the Am-241 source, selected mono-energetic beams at A2, and the K-edge energies of Ho, Yb and Re. The expected gain for each system was calculated by using the measured gain of the corresponding detector system with silicon sensor (\cite{mm_gain} and \cite{K4}) and multiplying it by the ratio of the energy needed to create an electron hole pair in the respective material ($W_{Si}/W_{CdTe} = 3.62/4.43 = 0.817$).

The alpha particle data provides a measure of the actual charge collection efficiency of the sensor. Since measurements were taken as a function of voltage, we can extract the $\mu_h\tau_h$ product of the material for both systems at different temperatures. 
To determine the $\mu_h\tau_h$ product we fit the simplified Hecht equation \cite{hecht} to the data points; the equation is $Q=Q_0 \left[ \frac{\mu_h\tau_h V}{d^2} \left( 1-\exp\left(-\frac{d^2}{\mu_h\tau_h V}\right) \right) \right]$, with $Q$ being the measured charge, $Q_0$ the deposited charge, $d$ the detector thickness and $V$ the applied bias. The individual results for -30~C, 0~C, and 20~C are (0.63$\pm$0.09)~$\times$~10$^{-4}$~cm$^2$/V, (0.63$\pm$0.10)~$\times$~10$^{-4}$~cm$^2$/V, and (0.77$\pm$0.13)~$\times$~10$^{-4}$~cm$^2$/V, respectively.
The average value of all measurements is $\mu_h\tau_h$~=~(0.68$\pm$0.11)~$\times$~10$^{-4}$~cm$^2$/V which is close to the literature values \cite{greiffenberg}, but smaller than the value of 88~cm$^2$/Vs~$\times$~2~$\upmu$s~=~1.76~$\times$~10$^{-4}$~cm$^2$/V used in our simulations. Within the experimental error, the results obtained from the alpha particle data do not support a temperature dependence of the  $\mu_h\tau_h$ product.

%

\begin{figure*}[tb!]
  \centering
  \begin{subfigure}[t]{0.45\textwidth}
	\includegraphics[width=\textwidth]{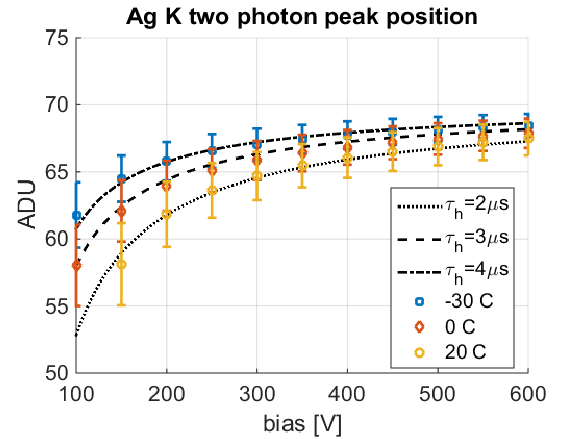}
	\caption{Effect of bias voltage on measured gain at different temperatures. The error bar on each data point indicates the width of the underlying distribution. Simulations are indicated by lines.}
  \end{subfigure}
\quad
  \begin{subfigure}[t]{0.45\textwidth}
	\includegraphics[width=\textwidth]{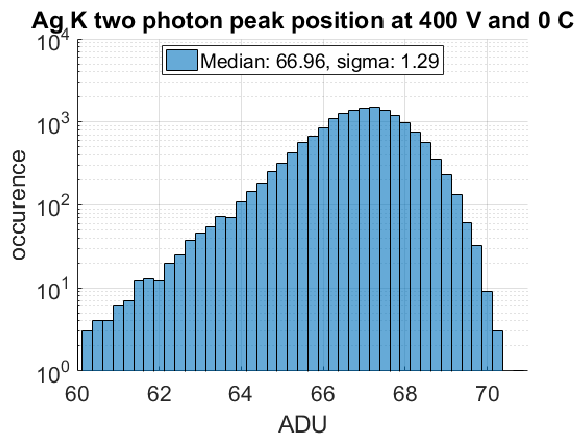}
	\caption{Histogram of the gain distribution for 400~V bias at 0~C. The skew of the distribution is possibly a result of the internal features of the CdTe crystal}
  \end{subfigure}
  \caption{Position of the Ag K line as a function of voltage and temperature measured with the MM-PAD detector.}
  \label{ag_kb}
\end{figure*}

To estimate the influence of voltage and temperature on the sensor response to photons we acquired several low flux data sets with the Ag tube and evaluated the position of the two photon Ag K line in the spectrum as a function of voltage and temperature. Without the use of a monochromator Ag K$_\alpha$ and Ag K$_\beta$ lines could not be separated and a weighted average energy was used instead. Additionally, looking at the position of the two photon peak is preferred over looking at the single photon peak, as the energy resolution of the MM-PAD system is not sufficient to resolve the overlap of the single photon Ag K line and the Cd and Te fluorescence lines. The position of the peak was determined by a fit to the data, providing more precise values than picking the ADU bin with the most counts. The results are presented in Figure \ref{ag_kb} and show a clear dependence on voltage and temperature. This dependence can be understood in terms of drift and trapping times in the detector. At higher voltages the detrapping of carriers is enhanced and the drift time is shorter,\footnote{The drift speed is the product of mobility and local electric field, which in turn is a function of the applied bias.} therefore fewer carriers are trapped and more charge is collected, leading to the observed higher peak position for higher voltages. 

The trapping time of charge carriers appears to have increased at lower temperatures, corresponding to a lower charge loss at lower temperature. Since the mobility of holes, which determine the actual drift time, remains almost constant in the investigated temperature regime \cite{mob}, the observed behavior is attributed to changes in the effective trapping time. Temperature dependence of the trapping time is reasonable and consistent with the known behavior of CdTe bulk semiconductors.\footnote{Trapping probability can be expressed as $1/\tau = \sigma N v_{th}$, with $\sigma$ being the capture cross section, $N$ being the trap density and $v_{th}$ being the thermal velocity of the charge carrier. Thus at higher temperatures we have higher thermal velocities and therefore lower trapping times (higher trapping probabilities).} Simulations of the expected peak value for hole trapping times $\tau_h$ of 2, 3 and 4~$\upmu$s have been included in Figure \ref{ag_kb}. 

The results from measurements using the Ag K line and alpha particles seem to contradict each other. The factor of 2 change in trapping time when going from -30~C to 20~C at constant mobility, as indicated by the photon measurement, cannot be supported within the experimental error of the measurement of a constant $\mu_h\tau_h$ product from alpha measurements for the same temperature range.

We note that alpha particles tend to deposit most of their energy close to the surface of the detector, while x-rays penetrate some distance into it before converting, resulting in a different average drift time. However this effect was included into the simulations and can be ruled out as a cause for the apparent contradiction. 

A possible explanation might be the densely ionizing nature of alpha particles. Densely ionized regions create local field distortions due to the plasma effect \cite{p1, p2, p3}. These distortions, in conjunction with the fact that (filled) local traps might be acting as recombination centers, could in principle influence the measured effective trapping time and lead to different results for the two measurement methods.

\subsection{Edge response}

\begin{figure*}[tb!]
  \centering
  \begin{subfigure}[t]{0.49\textwidth}
	\includegraphics[width=\textwidth]{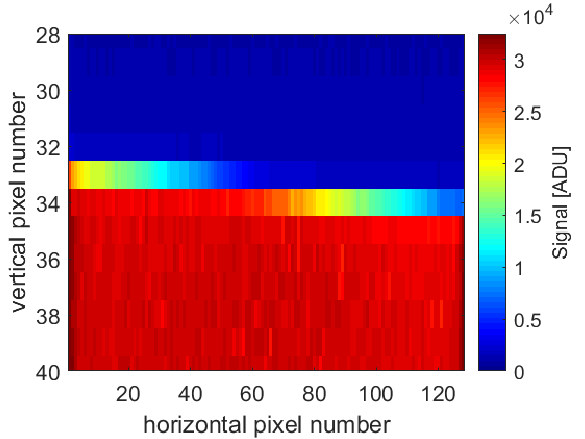}
	\caption{Typical image of the knife edge. Note the image is expanded vertically (non-square aspect ratio).}
  \end{subfigure}
\quad
  \begin{subfigure}[t]{0.45\textwidth}
	\includegraphics[width=\textwidth]{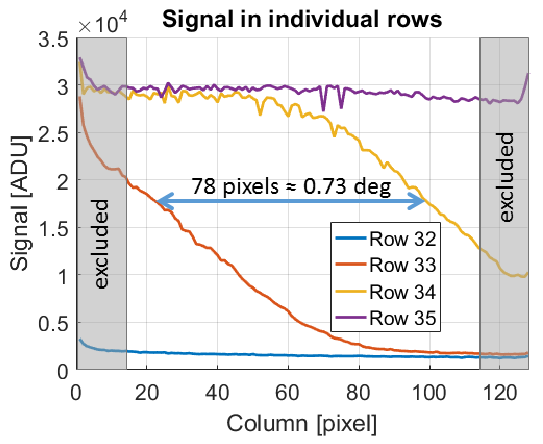}
	\caption{Typical response of 4 consecutive rows located around the edge.}
  \end{subfigure}
  \centering
	\caption{Data of rows of pixels of the MM-PAD located around the knife edge illumination. The signal shown is an average of 100 individual frames. The gray regions in b) indicate the areas excluded from the analysis due to possible rim effects.}
	\label{rows}
\end{figure*}


To determine the response of a pixel to photons as a function of incident position we measured the edge spread function (ESF) using the Ag x-ray tube at 47~kV.
A tungsten knife edge was mounted in front of the detector surface with an intentionally small angular misalignment with respect to a row of pixels; in our case the tilt is $\theta = 0.73$ degrees. 
In this way we get several rows of pixels where the sensitive area of a pixel is partially covered and we can transform the column coordinate to an effective displacement $d_{disp}=n p \sin \theta$, with $n$ being the column coordinate and $p$ being the pixel pitch of 150~$\upmu$m. An example of the signal in four consecutive rows is shown in Figure \ref{rows}. 

As the angle is large enough that least two rows are partially covered, the tilt of the knife edge could be verified from the data and the rows can be shifted and their values averaged to reduce the noise in the data.

Since the ideal edge spread function is the system response to a step function, the derivative of the ESF corresponds to the system point spread function (PSF). In general, derivatives of measurement data are very sensitive to measurement noise. 
Although averaging of frames and adjusted rows reduced the noise significantly, additional staged median and running average filters were used to produce smoother derivatives. The use of these filters reduces spatial resolution and as a consequence, the spatial resolution here is limited to approximately 10~$\upmu$m precision.

\begin{figure*}[tb!]
  \centering
  \begin{subfigure}[t]{0.45\textwidth}
	\includegraphics[width=\textwidth]{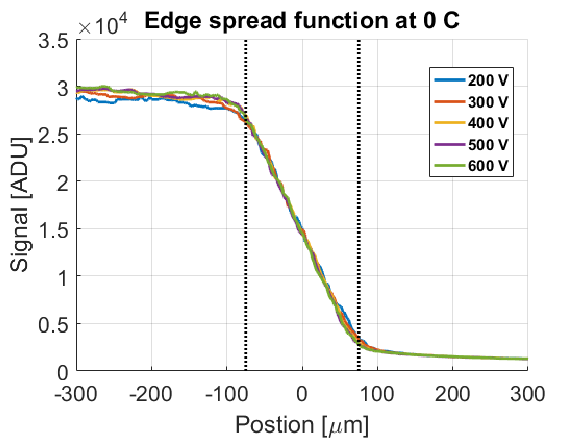}
	\caption{Edge spread function at different bias voltages for a fixed temperature.}
	\label{ESF}
  \end{subfigure}
\quad
  \begin{subfigure}[t]{0.45\textwidth}
	\includegraphics[width=\textwidth]{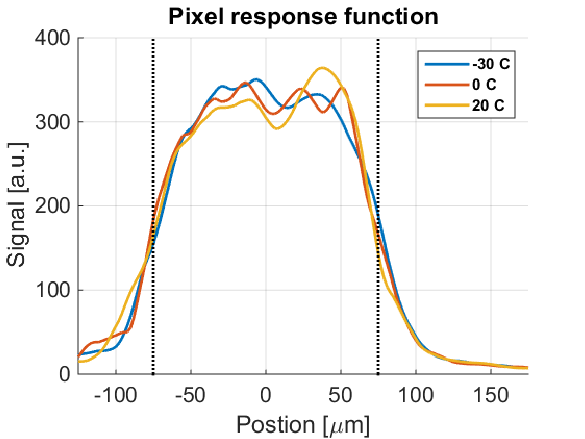}
	\caption{Pixel response function calculated from the voltage averaged ESF at different temperatures.}
	\label{PRF}
  \end{subfigure}
  \caption{Edge response measured with the MM-PAD detector. The dashed vertical lines indicate the pixel size for pixel centered at zero.}
  \label{edge}
\end{figure*}

The reconstructed edge spread function and the pixel response function of the system are shown in Figure \ref{edge}. It is evident from Figure \ref{ESF} that the edge response depends very little on the applied voltage, and, within the experimental precision, we could not determine any influence of the temperature on the pixel response function either, as shown in Figure \ref{PRF}. After filtering the residual noise in the pixel response function is about $\pm$~10\%, as evidenced by the variability of the flat-top region. Despite the noise one can estimate from the rise and fall of the pixel response function that for illumination from the x-ray tube the average charge cloud size seen at the readout electrodes is approximately 50~$\upmu$m.

Since we were using a filtered x-ray tube spectrum with components above and below the K-edges of Cd and Te to measure the ESF the measured charge cloud size of approximately 50~$\upmu$m is an effective size that includes the spreading effects of the fluorescence and other second order effects. This is supported by the lack of change in size for different biases. If the effective cloud size was dominated by the lateral spread due to diffusion or mutual electrostatic repulsion of the charge carriers this would produce a measurable effect in our data. Measurements at acceleration voltages below the K-edges of Cd and Te (not shown) reveal a decrease of the long tails (caused by fluorescence parallel to the detector plane) in both ESF and PSF, but within our experimental precision we could not observe a significant change in the size of the charge cloud. This indicates that the overall contribution of the fluorescence photons to the cloud size is small, which is expected from simulations \cite{david}.\footnote{Reference \cite{david} does not model our experimental situation exactly, but shows that at 80~keV photon energy, when compensating for charge shared events, only 18\% of all photons produce a fluorescence photon that escapes out of a 110~$\upmu$m pixel.}

\subsection{Polarization}

\begin{figure*}[tb!]
  \centering
  \begin{subfigure}[t]{0.45\textwidth}
	\includegraphics[width=\textwidth]{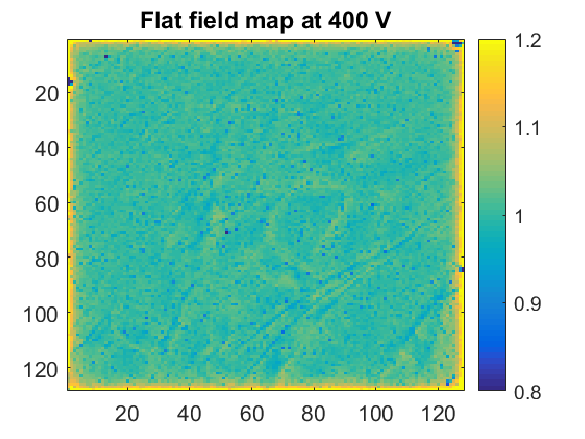}
	\caption{Before dosing.}
  \end{subfigure}
  \quad
  \begin{subfigure}[t]{0.45\textwidth}
	\includegraphics[width=\textwidth]{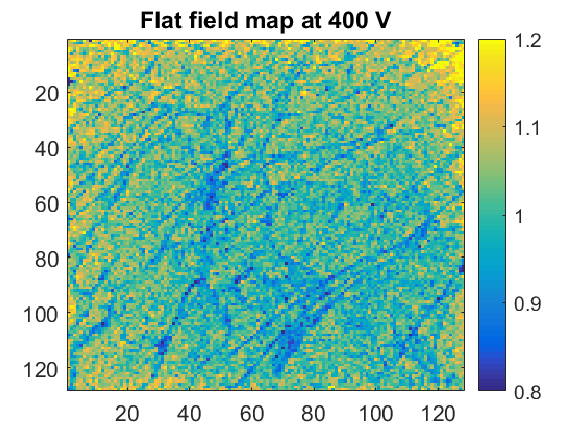}
	\caption{After exposure to 3~$\times$~10$^{11}$ photons/mm$^2$.}
  \end{subfigure}
	\caption{Effect of polarization for the MM-PAD detector at 0~C. Note the different color scale of the unpolarized flat field compared to Figure \ref{flat_field}.}
	\label{ff_pol}
\end{figure*}


The polarization of CdTe and its associated changes in current and, hence, offset correction depend on local material properties (e.g., the trap density). Therefore, changes in response are modulated locally. This is evidenced by the evolution of the flat field response of the detector as shown in Figure \ref{ff_pol}, which shows an increase in contrast for the network of lines that are a feature of the sensor. 

In the following we investigate three polarization effects in more detail: A) the reduction of signal (count rate deficit) with dose, B) the deterioration of the image uniformity with time and C) the lateral displacement of signal due to differential dosing.

It should also be noted that the increased signal towards the edge of the sensor, the so-called rim-effect, also is reduced with increased polarization. For the time being, we can only speculate about the mechanisms involved in this behavior, but in general it is assumed than rim effects originate from the distortions in the local drift field due to the proximity of the sensor's cut edge. It is plausible that charge build up from trapped charges (i.e., polarization) can influence the amount of distortion to which a given pixel is subject.

\subsubsection{Signal reduction}

The response of a CdTe sensor is reduced after it has been exposed to a significant number of photons. In our tests, dosing of the sensor was achieved by placing it very close ($\approx$~5~cm) to the output of the silver tube, biased to 47~kV, keeping only the 1~mm aluminum filter. In this way we achieved an approximate dose rate of 11~Gy/min or 3~$\times$~10$^7$ photons/mm$^2$/s. The dose rate was measured using the detector itself and the calibration presented above. The conversion to photons is only approximate as it used a generic Ag anode x-ray tube spectrum.

Please note that our detector systems are charge integrating and therefore do not count individual pulses from incoming photons in the same way as photon counting detectors. Nevertheless we observed a reduction of measured signal with exposure that is similar to what is commonly referred to as count rate deficit when investigated with high photon fluxes and counting detectors. 

\begin{figure*}[tb!]
  \centering
  \begin{subfigure}[t]{0.45\textwidth}
	\includegraphics[width=\textwidth]{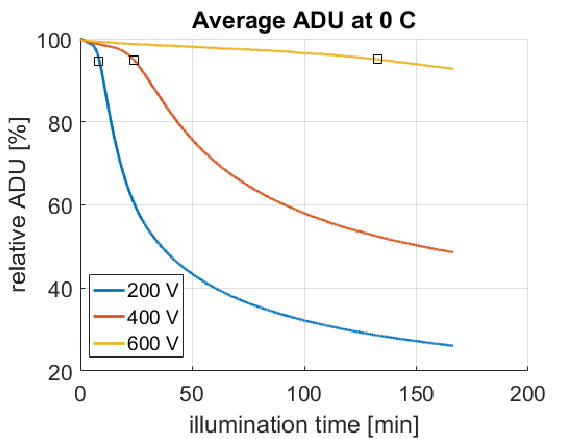}
	\caption{Decrease of normalized signal with time for different voltages.}
	\label{crd1}
  \end{subfigure}
\quad
  \begin{subfigure}[t]{0.45\textwidth}
	\includegraphics[width=\textwidth]{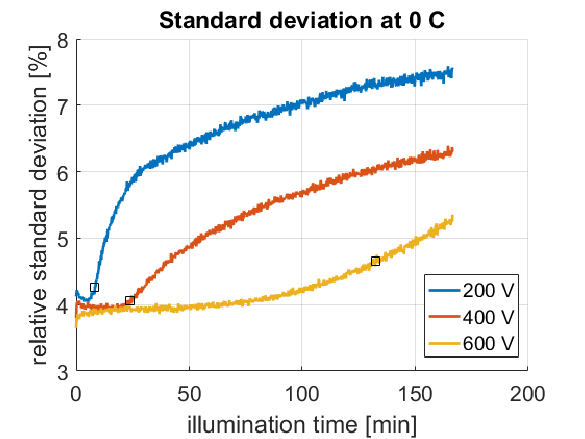}
	\caption{Relative standard deviation in flat field images as a function of time for different voltages.}
	\label{crd2}
  \end{subfigure}

  \caption{Effect of exposure of the MM-PAD with 3$\times$10$^7$ ph/mm$^2$/s for the Al filtered Ag tube. The polarization time, $t_{pol}$, is indicated by the black square on each curve.}
  \label{crd}
\end{figure*}

Using the x-ray tube as a source, we have studied this effect of reduced signal as a function of sensor bias and time. Figure \ref{crd1} shows a slow decrease of the signal to approximately 95\% of the initial value followed by a steep decrease followed by a slower decline. We call the time to reach 95\% of the initial signal the polarization time, $t_{pol}$. After the system starts to polarize ($t>t_{pol}$) the network of lines in the flat field increase in contrast, as evidenced by the increased standard deviation in flat field images shown in Figure \ref{crd2}. 

Increasing the bias voltage from 200~V to 600~V increased the polarization time at this x-ray flux from under 10 minutes to over 2 hours, underlining the importance of operating CdTe sensors at high bias voltages.

\begin{figure*}[tb!]
  \centering
  \begin{subfigure}[t]{0.45\textwidth}
	\includegraphics[width=\textwidth]{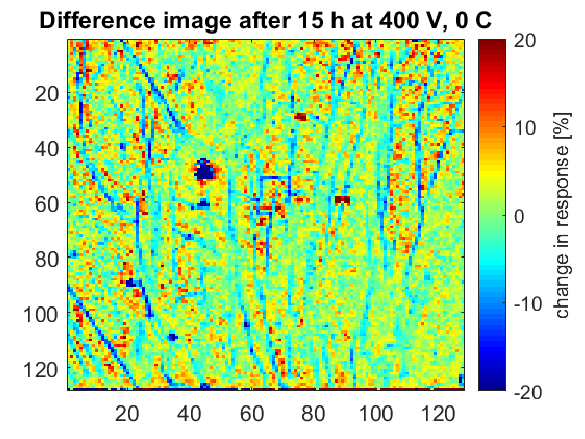}
	\caption{Relative difference in response of the system after long operation at 400~V and 0~C by dividing an image taken after 15 hours by the very first image taken. }
	\label{diff}
  \end{subfigure}
\quad
  \begin{subfigure}[t]{0.45\textwidth}
	\includegraphics[width=\textwidth]{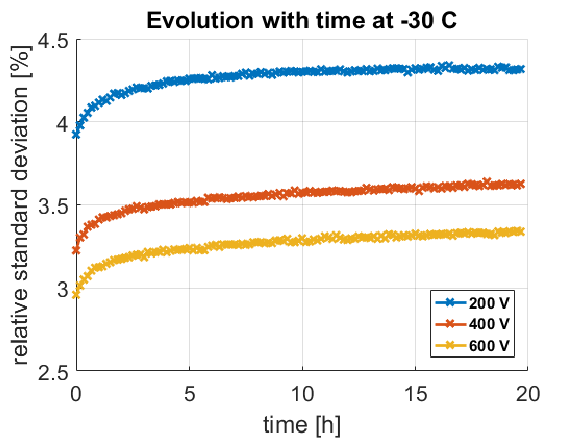}
	\caption{Relative standard deviation in images as a function of time for different voltages. At -30~C the image deteriorates slowly, with a slightly faster deterioration in the first few hours.}
	\label{d1}
  \end{subfigure}
%
\smallskip

  \begin{subfigure}[t]{0.45\textwidth}
	\includegraphics[width=\textwidth]{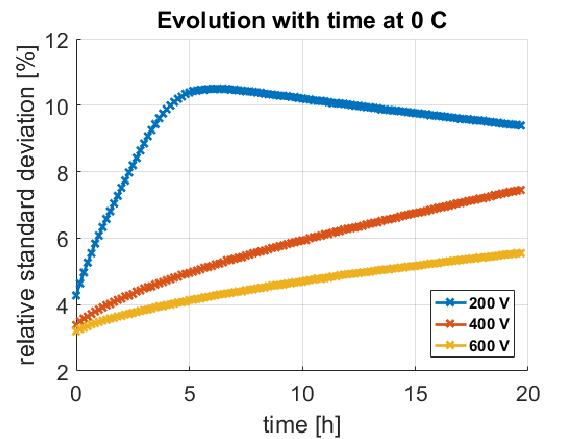}
	\caption{Relative standard deviation in images as a function of time for different voltages. At 200~V a peak distortion occurred after approximately 5 hours of operation, followed by a slow improvement in image uniformity. The response for the other voltages follows a behavior similar to that for -30 C, but at an accelerated rate. }
	\label{d2}
  \end{subfigure}
\quad
  \begin{subfigure}[t]{0.45\textwidth}
	\includegraphics[width=\textwidth]{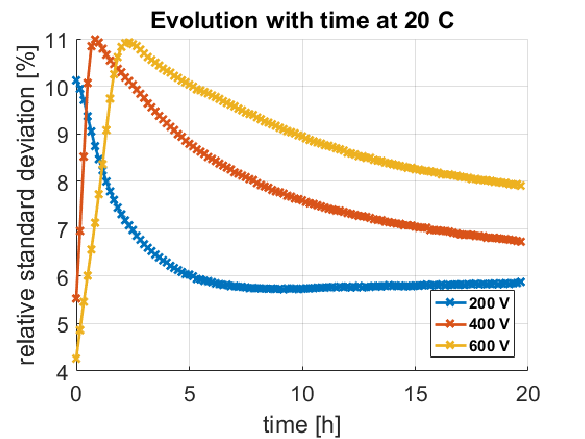}
	\caption{Relative standard deviation in images as a function of time for different voltages. All curves are characterized by a quick rise to a peak value and a slow decay to a saturation value. At 200~V the peak distortion occurred before the 3 minute wait time after the application of the bias had ended.}
	\label{d3}
  \end{subfigure}
  \caption{Signal deterioration with time. The large dark spot at approximately (45,50) is related to a surface defect introduced in the assembly process and has been excluded from the analysis. For this measurement a Keck PAD system was used, so the observed network of lines is different from the one shown before, as the sensor was made from a different CdTe wafer.}
  \label{dwt}
\end{figure*}

Data taken at different temperatures revealed a temperature dependence of the polarization time. The decrease in signal with time remained qualitatively similar at different temperatures, but with a sharper `knee' and steeper slope at lower temperature compared to a more gradual transition at higher temperatures.


After each voltage step the sensor was reset with the standard cycle. Inspection of the background corrections (an average of 100 dark frames in this case) before dosing and after the reset showed negligible differences on the order of less than 1 ADU per pixel. It is also observed that the reset is sufficient to restore the signal to its original value with an error of less than 0.5\%.

\subsubsection{Response deterioration with time}

\begin{figure*}[tb!]
  \centering
  \begin{subfigure}[t]{0.45\textwidth}
	\includegraphics[width=\textwidth]{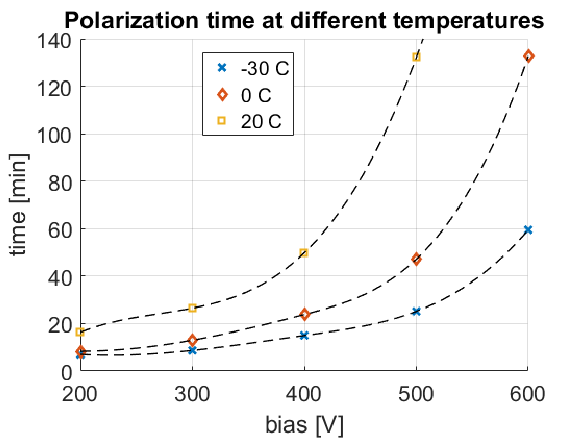}
	\caption{Exposure time after which the signal level has dropped to 95\% of its original value. Higher values indicate better performance. The black dashed lines are cubic spine interpolations of the data points to guide the eye. Data taken with the MM-PAD. }
	\label{t1}
  \end{subfigure}
\quad
  \begin{subfigure}[t]{0.45\textwidth}
	\includegraphics[width=\textwidth]{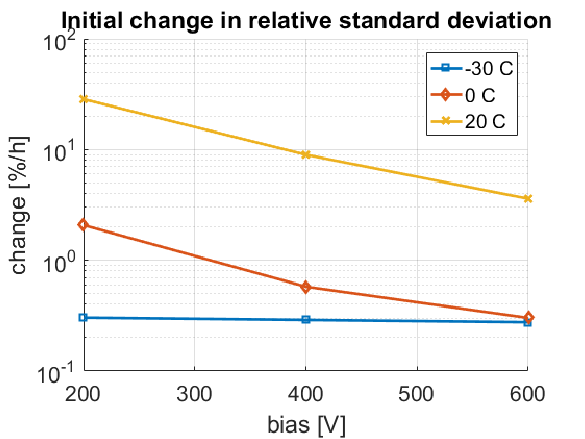}
	\caption{Initial slope of the deterioration with time. Lower values indicate better performance. The curve at -30 C reproduces the same trend as the other two curves albeit at a much smaller slope. The data points are connected by straight lines to guide the eye. Data taken with the Keck PAD.}
	\label{t2}
  \end{subfigure}
  \caption{Figures of merit for dose dependent and time dependent polarization as a function of temperature and bias.}
  \label{t_pol}
\end{figure*}

The response of the CdTe sensor to radiation does not only change as a function of exposure, as explained above, but also as a function of time, in the absence of accumulated x-ray dose. 

To investigate this behavior, we took a flood illumination of the sensor every 10 minutes and looked at the changes in the response over time. In order to avoid polarizing the sensor by the exposure, each illumination was kept short. The total illumination time was measured to be less than 2.4~s per data point, which led to a `on' duty cycle of 0.4\% and a total exposure to less than 1~$\times$~10$^{10}$ ph/mm$^2$ after 20 hours.

While the sensor response to exposure shows clear features of a count rate deficit, as outlined previously, during this measurement the signal stayed approximately constant with time, but showed an increase in the relative standard deviation due to a reduction in image uniformity. Variance in the incoming photon flux can be excluded as the source for this, as the tube intensity was monitored with a silicon sensor next to the device under test.

\begin{figure*}[tb!]
  \centering
  \begin{subfigure}[c]{0.45\textwidth}
	\includegraphics[width=\textwidth]{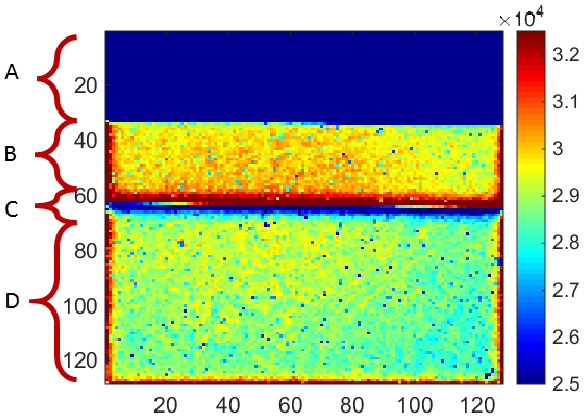}
	\caption{2D image showing the distortions. The region A)~was shielded, B)~was shielded during the first exposure, C)~is the region of displacement artifacts, and D)~was exposed to the full dose.}
	\label{2d}
  \end{subfigure}  
  \quad
  \begin{subfigure}[c]{0.45\textwidth}
	\includegraphics[width=\textwidth]{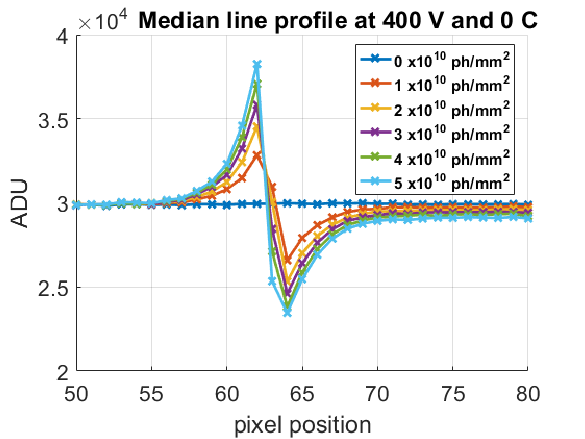}
	\caption{Median of all line cuts along the vertical axis (top to bottom in the image to the left) for different exposures.}
	\label{line_cut}
  \end{subfigure}  

\bigskip

   \begin{subfigure}[t]{0.45\textwidth}
	\includegraphics[width=\textwidth]{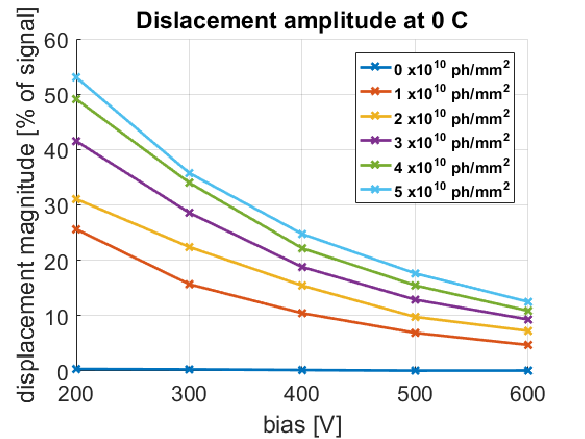}
	\caption{Amplitude of the bipolar lateral displacement effect as a function of exposure and applied bias for a fixed temperature.}
	\label{amp_v}
  \end{subfigure}  
  \quad
  \begin{subfigure}[t]{0.45\textwidth}
	\includegraphics[width=\textwidth]{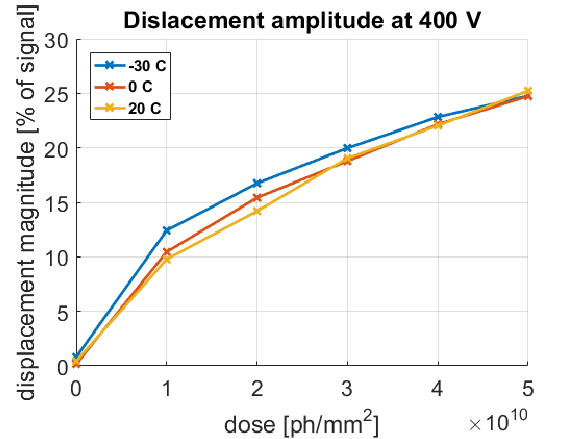}
	\caption{Temperature dependence of the displacement amplitude for a fixed voltage.}
	\label{amp_t}
  \end{subfigure}
  \caption{Response of the MM-PAD system after non-uniform dosing using a knife edge with up to 5~$\times$~10$^{10}$ ph/mm$^2$}
  \label{lines}
\end{figure*}

Figure \ref{diff} shows the difference in response of the system after long operation at 400~V and 0~C by dividing an image taken after 15 hours by the very first image taken. We observe that certain hot ($>+ 20 \%$ signal) and cold ($< - 20 \%$ signal) spots have formed and the network of lines already known from the flat field images shown in Figure \ref{flat_field} introduced a modulation on the order of $\pm$ 10 \%, with some modulation amplitudes at more than double that value.\footnote{The network of lines in Figures \ref{flat_field} and \ref{diff} do not match, as sensors from different CdTe wafers were used for these measurements.}

When looking at the distortion of the images as a function of time for different biases and temperatures (Figures \ref{d1}, \ref{d2}, \ref{d3}), we noticed a seemingly universal behavior, namely the continuous increase in distortion until a peak distortion is reached, followed by an improvement of the image uniformity until a stable saturation value is reached. The time constants of this rise and decay in non-uniformity are influenced by temperature and the applied bias. We observed that lower temperatures and higher biases increased the time constants and thereby reduced the amount of additional non-uniformity introduced per unit time for typical operating conditions.

Lastly, the characteristic figures of merit of the system for different biases and temperatures are shown in Figure \ref{t_pol}. Note that the temperature dependence of the two figures of merit are opposed. For this reason our standard operating temperature of 0~C is a compromise to reduce the influence of both effects.

\subsubsection{Lateral displacement}


\begin{figure*}[p!]
  \centering
%
  \includegraphics[width=1.0\textwidth]{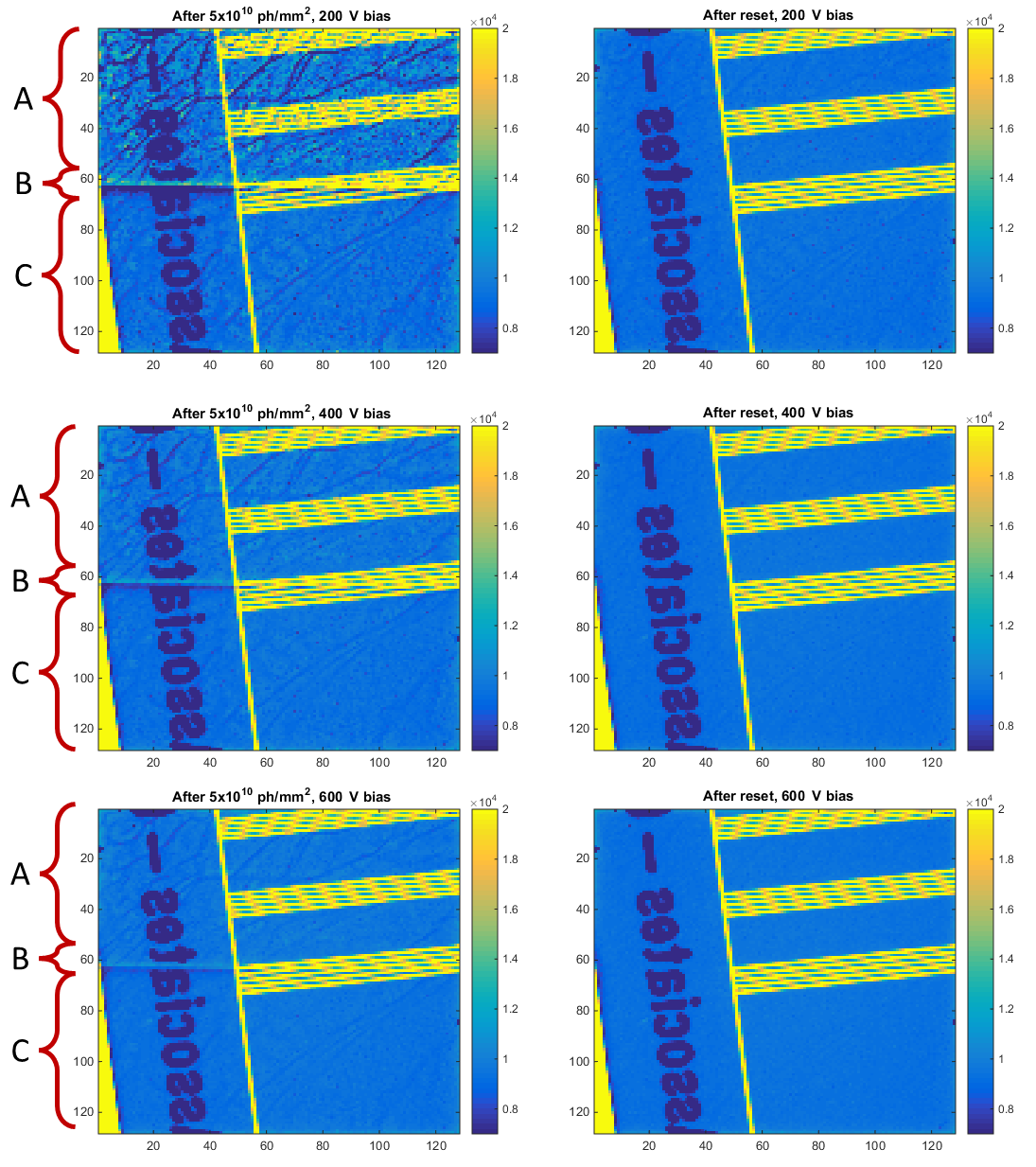}
  \caption{Response of the MM-PAD system to a short exposure after dosing the lower half of the sensor with 5~$\times$~10$^{10}$ ph/mm$^2$ at 600~V bias. The response before and after the reset is shown in the left and right columns, respectively. Region A was undosed and shows increased variance but the same average signal. Region B is in between the dosed and undosed region and shows the lateral displacement artifact. Region C was dosed and shows a reduction in signal, but less variance than in region A. Resetting the detector removed the effects in all three regions.}
  \label{partial}
\end{figure*}

Image degradation occurs after heavy dosing. This is presumed to be due to the presence of trapped charges that locally distort the electric bias field and push signal collection away from a straight line path perpendicular to the sensor faces.

Differential dosing, i.e., dosing only parts of the sensor, can create additional local distortions in the recorded images, as shown in Figure \ref{2d}, where the lower part of the image has been dosed while the upper part was shielded. The shown image was taken after moving the illuminated area up, thereby revealing a bipolar distortion artifact across the illumination boundary. Line cuts (i.e., graphs of recorded intensity along a line of pixels) over the distortion are shown in Figure \ref{line_cut} as a function of exposure. The magnitude of this effect increases with increased dose, decreases with increased bias, and is almost independent of temperature, as Figures \ref{amp_v} and \ref{amp_t} show. Note that the displacement effect appears at one order of magnitude lower dose compared to the signal reduction due to dose.

This effect can be difficult to see in non-uniform illuminations which are common in, e.g., crystallography. It could be seen as an increase in the width or a lateral displacement of a Bragg spot or powder diffraction ring or as `ghosting' of a previous pattern in subsequent images. 

Figure \ref{partial} shows the practical implications of all three investigated polarization effects on a line pair mask. While the undosed region retained a higher average signal, it also exhibited less uniformity than the dosed region, and signal was lost in places coinciding with the network of lines.
A standard reset cycle restored the original conditions, removed the ghosting effect and recovered the original signal level.

\begin{figure*}[tb!]
  \centering
  \includegraphics[width=1.0\textwidth]{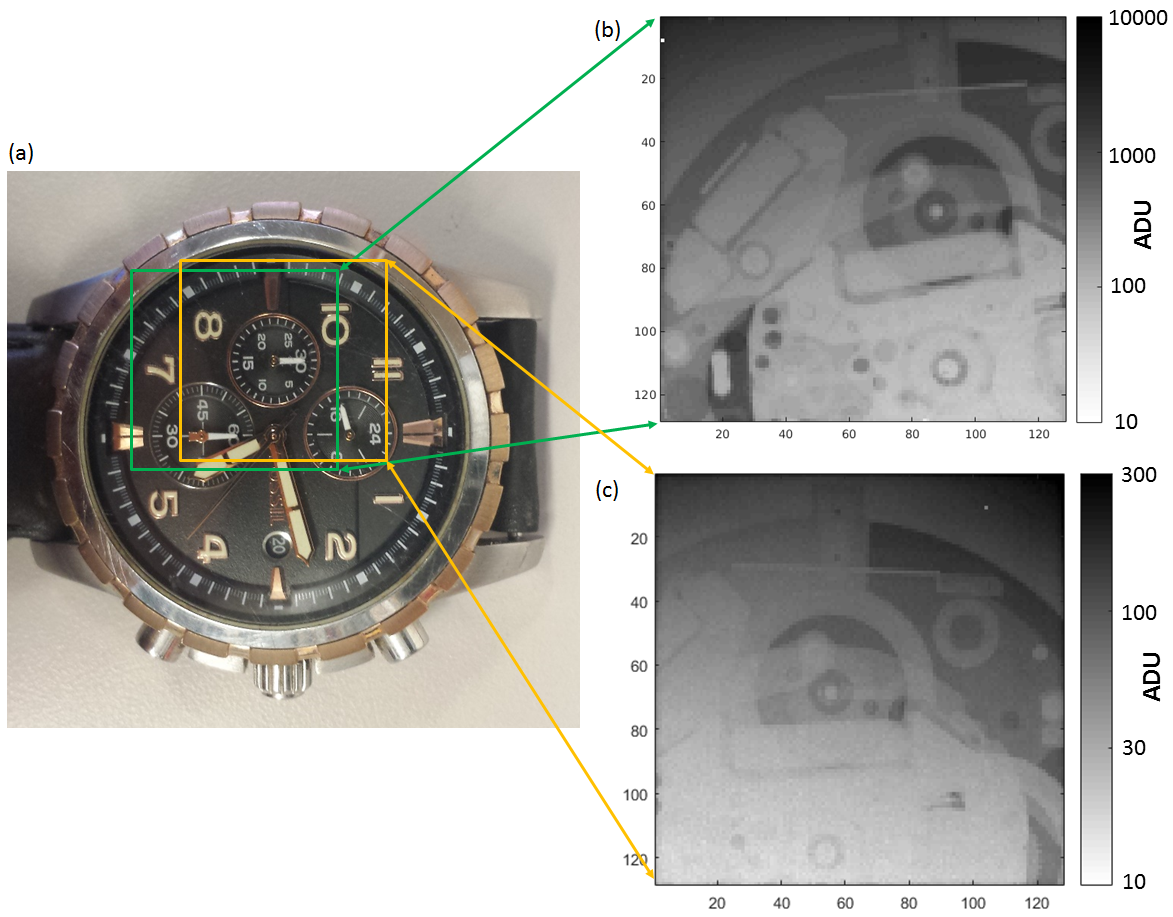}
  \caption{Visible light image (a) and corresponding radiographs of a watch acquired with MM-PAD systems with either CdTe (b) or silicon sensors (c). The upper radiograph is the average of 1000 individual frames of with 1~ms illumination time each. The lower radiograph is an average of 1000 individual frames with 500~ms illumination time each. Note the logarithmic gray scale spans 3 orders of magnitude from less than 1 photon per frame (10~ADU) to more than that 300 photons/frame (10000~ADU) for the upper image (CdTe) and a factor of 30 for the lower image (silicon). The image taken with the CdTe sensor allows one to discern more detail in the central part of the watch.}
  \label{watch}
\end{figure*}

\section{Results}

\begin{figure*}[tb!]
  \centering
  \begin{subfigure}[t]{0.45\textwidth}
	\includegraphics[width=\textwidth]{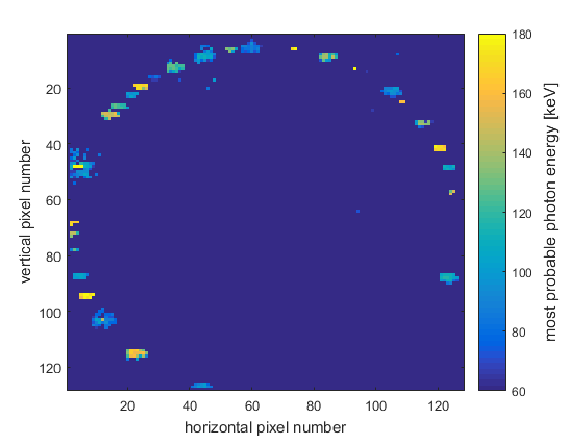}
	\caption{Most probable energy of the diffracted photons for each pixel. The photon energy was determined from the central position of the first mode in each pixel's data.}
  \end{subfigure}
\quad
  \begin{subfigure}[t]{0.45\textwidth}
	\includegraphics[width=\textwidth]{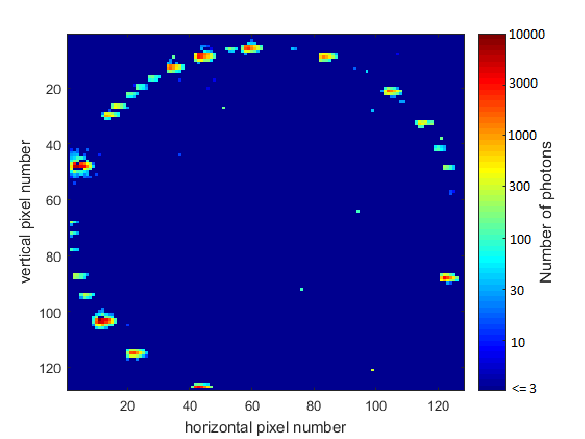}
	\caption{Number of photons registered in each pixel displayed on a logarithmic color scale. The number of photons was derived from the sum of events in each mode multiplied by the mode number.}
  \end{subfigure}
  \caption{Experimental results from the Keck detector with CdTe for GGG in Laue geometry exposed with a white beam in a low flux scenario. The whole data set consists of about 8000 individual frames.}
  \label{ggg}
\end{figure*}

\begin{figure*}[tb!]
  \centering
	\includegraphics[width=0.9\textwidth]{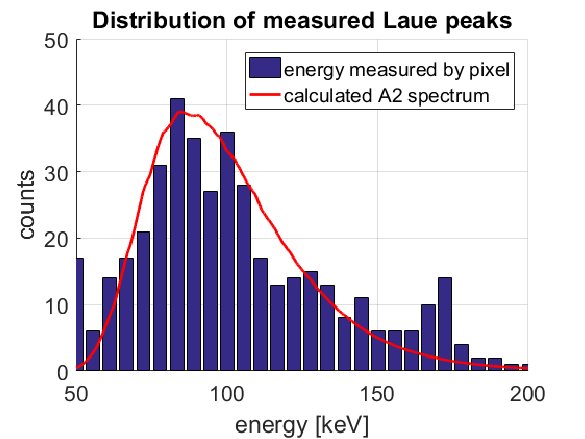}
  \caption{Distribution of the energy determined by the pixels of each Laue peak. The distribution coincides roughly with the beam spectrum at A2 and diffraction peaks with energies as high as 200~keV have been detected.}
  \label{ggg_spectrum}
\end{figure*}


The usefulness of the detector as an imaging system was tested using an MM-PAD assembly to image an object in transmission. A wristwatch was mounted in front of the detector with the x-ray tube approximately 1~m away, resulting in negligible magnification. The resulting transmission radiograph is shown in Figure \ref{watch} together with a photo of the watch taken by visible light and a radiograph with an MM-PAD with a silicon sensor. 

The watch is a macroscopic object containing many thick parts. This means that only the high energy part of the x-ray tube spectrum will penetrate through it, making the high quantum efficiency of, e.g., our CdTe sensors mandatory for reasonable imaging. Figure \ref{watch} shows internal features of the watch in higher detail and contrast than seen in the silicon version. Presumably the contrast of the image taken with the silicon sensor is reduced due to Compton scattering of high energy photons within the silicon sensor. Note that for this comparison the silicon sensor was exposed much longer than the CdTe sensor to reach a roughly comparable image quality in both images. 

The Keck PAD system was used in a white beam experiment at the A2 beamline at CHESS. A Gadolinium-Gallium-Garnet (GGG) crystal was the sample and both sample and detector were aligned such that a circle of Laue reflections was visible on the detector (see Figure \ref{ggg}), although some of it was blocked by a beam stop in the lower right hand corner of the images.

An integrating detector is only sensitive to the signal generated by the deposited energy, which is approximately the sum of all photons multiplied by their respective energies. Thus, in a monochromatic experimental situation, which is not the case here, the signal would be proportional to the number of photons. Using the A2 white beam for Laue diffraction, we could not derive the number of photons directly from the signal and had to use another method.

By lowering the incoming photon flux such that individual frames were sparsely populated we could look at the photon spectrum on a per pixel basis. From the spectra we could then determine the energy of the photon peak and number of photons separately for each pixel.\footnote{We defined an energy threshold per pixel at half the photon peak energy, $E$. Events above threshold were counted as one photon each, events above 3 times the threshold ($1.5E$) were counted as two photons, and so on.} Figure \ref{ggg} shows the results of this method: an energy map (showing the central position of the photon peak) and a photon distribution map (showing the total number of photons per pixel in the data). 

The energy distribution of the Laue peaks is shown in Figure \ref{ggg_spectrum}, and compared to the calculated spectrum of the A2 beamline. The figure clearly shows that Laue peaks all the way to the high energy tail of the spectrum at approximately 200~keV were detected.


\section{Conclusions and next steps}

CdTe of 750 $\upmu$m thickness expands the usable energy range for detector systems from about 25~keV ($\approx$20\% QE with a 500~$\upmu$m silicon sensor) to about 150~keV ($\approx$20\% QE with a 750~$\upmu$m CdTe sensor), with $>$90\% quantum efficiency to about 65~keV. 


Our characterization of CdTe sensors bonded to our detectors has led us to the conclusion that the material is suitable for use in a broad range of experimental applications. Additionally we found the performance in hole-collecting mode better than originally anticipated and comparable to other CdTe-based detectors with Schottky contacts.\footnote{Electron collection is commonly considered preferable over hole collection due to the larger $\mu\tau$ product for electrons.} Most of the effects described in this work amount to relatively small scale distortions of the measured signal. However, as shown herein, an understanding of these effects is important in order to understand the limitations of measurements taken with the detectors. We note that both CdTe-bonded Keck PAD and MM-PAD units have already been used in a number of experiments at CHESS, and further experiments are planned at both CHESS and the Advanced Photon Source. 

The tested modules show very few material defects and excellent uniformity in the response to x-rays. Whether this was caused by an increased crystal quality, improvements in processing, the fact that we collected holes instead of electrons, the fact that we used an integrating system instead of a counting system, or a combination of any or all of these factors remains unresolved at this point.

The performance of our modules was characterized as a function of dose and time and known polarization effects could be reproduced and observed. The system response was robust up to a certain dose, at which point there was a rapid decline in response followed by a continued but slower reduction in signal. The system response in time was characterized by a gradual increase in inherent sensor fluctuations. We found that the temperature dependence for the signal reduction and the temperature dependence of the increase of inherent fluctuations with time were opposed. Therefore, we chose a standard operating temperature of 0~C as a compromise.

Further, we found that the lateral displacement after differential dosing was not a permanent feature in our Schottky type sensors, in contrast to the observations of other authors for ohmic type sensors \cite{grains1}. The reasons for this are unclear, but the different sensor design and our more aggressive reset scheme might contribute to this finding. The effect of the lateral displacement was already measurable at one order of magnitude less dose than the signal reduction with dose. 

A reset procedure effective at removing the effects of polarization described here was developed and employed.


Being designed for fast imaging experiments, the Keck PAD system allows investigations on timescales $<$~1~$\upmu$s. In fact, integrating pixels are essentially high speed, spatially resolving, low noise electrometers. This makes our detectors well suited tools to study the trapping and detrapping dynamics in CdTe. Further studies on this topic are currently underway.

\acknowledgments
The authors would like to thank Dominic Greiffenberg from the Paul Scherrer Institute in Switzerland for valuable discussions and input, Wataru Inui at Acrorad, Co., Ltd for his contributions to the sensor design and production, and Klaus Harkonen, Limin Lin, and Konstantinos Spartiotis at Oy Ayad Ltd for the hybridization of the assemblies.

This research is supported by the U.S. National Science Foundation and the U.S. National Institutes of Health/National Institute of General Medical Sciences via NSF award DMR-1332208, U.S. Department of Energy Award DE-FG02-1 0ER46693 and DE-SC0016035, and the W. M. Keck Foundation. The MM-PAD concept was developed collaboratively by our detector group at Cornell University and Area Detector Systems Corporation (ADSC) in Poway, CA, USA.



\nocite{*}
\bibliographystyle{aipnum-cp}%

\end{document}